\documentclass[pdftex,twocolumn,epjc3]{svjour3}          

\RequirePackage[T1]{fontenc}

\smartqed  

\RequirePackage{graphicx}
\usepackage{latexsym}
\usepackage{amssymb, amsmath, } 
\usepackage{subcaption}
\RequirePackage{mathptmx}      
\RequirePackage{flushend}
\RequirePackage[numbers,sort&compress]{natbib}
\RequirePackage[colorlinks,citecolor=blue,urlcolor=blue,linkcolor=blue]{hyperref}

\journalname{Eur. Phys. J. C}

\begin{document}

\title{White dwarfs with a surface electrical charge distribution: Equilibrium and stability}


\author{G. A. Carvalho\thanksref{e1,addr1}
        \and
        Jos\'e D. V. Arba\~nil\thanksref{addr4}
        \and
        R. M. Marinho Jr\thanksref{addr1}  
        \and
        M. Malheiro\thanksref{addr1}
}

\thankstext{e1}{e-mail: araujogc@ita.br}

\institute{Departamento de F\'isica, Instituto Tecnol\'ogico de Aeron\'autica, S\~ao Jos\'e dos Campos, SP, 12228-900, Brazil\label{addr1}
          \and
          Departamento de Ciencias, Universidad Privada del Norte,
Avenida Alfredo Mendiola 6062 Urbanizaci\'on Los Olivos, Lima, Peru\label{addr4}
}

\date{Received: date / Accepted: date}

\maketitle

\begin{abstract}
The equilibrium configuration and the radial stability of white dwarfs composed of charged perfect fluid are investigated. These cases are analyzed through the results obtained from the solution of the hydrostatic equilibrium equation. We regard that the fluid pressure and the fluid energy density follow the relation of a fully degenerate electron gas. For the electric charge distribution in the object, we consider that it is centralized only close to the white dwarfs' surfaces. We obtain larger and more massive white dwarfs when the total electric charge is increased. To appreciate the effects of the electric charge in the structure of the star, we found that it must be in the order of $10^{20}\,[{\rm C}]$ with which the electric field is about $10^{16}\,[{\rm V/cm}]$. For white dwarfs with electric fields close to the Schwinger limit,  we obtain masses around $2\,M_{\odot}$. We also found that in a system constituted by charged static equilibrium configurations, the maximum mass point found on it marks the onset of the instability. This indicates that the necessary and sufficient conditions to recognize regions constituted by stable and unstable equilibrium configurations against small radial perturbations are respectively $dM/d\rho_c>0$ and $dM/d\rho_c<0$.
\end{abstract}

\section{Introduction}
\label{intro}

In recently years, observations reveals about the existence of some peculiar super-luminous type Ia supernovae (SNIa) \cite{Howell2006,scalzo2010,Hicken2007,yamanaka2009}. These types of supernovae are of particular interest, since their explosions are the most predictable and, often, the brightest events in the sky. In some works is suggested that a possible progenitor of such a peculiar SNIa is the super-Chandrasekhar white dwarf, which exceeds sig\-ni\-fi\-can\-tly the standard Chandrasekhar mass limit $1.44M_{\odot}$. The mas\-ses estimated for these super-Chandrasekhar white dwarfs are between $2.1-2.8M_{\odot}$, see, e.g., \cite{taub2011,silverman2011}. 

In the General Relativity scopes, these particular objects have attracted the attention of several authors, who in different works, propose diverse models in order to explain these super-Chandrasekhar white dwarfs. For instance, we found works where are considering white dwarfs with a strong magnetic field \cite{das2012,dasPRL}, in rotation and with different topologies for magnetic field \cite{Boshkayev2013a, Franzon2015, Subramanian2015, Bera2016, Otoniel2016} and in the presence of an electrical charge distribution \cite{liu2014}.

In what concerns to magnetized white dwarfs, in \cite{das2012, dasPRL} are found that a uniform and very strong magnetic field can overcome the white dwarf maximum mass up to $\sim2.9\,M_{\odot}$. Albeit these objects exceed significantly the Chandrasekhar mass limit, they suffer from severe stability issues as discussed in the literature \cite{chamel2013,dongPRL,jaziel2014,nityananda2015,carvalho2018}. In \cite{Boshkayev2013a, Franzon2015, Subramanian2015, Bera2016, Otoniel2016} are found that white dwarfs with rotation and different topology for the magnetic field could reach masses up to $5M_{\odot}$.

In electrically charged white dwarf \cite{liu2014}, Liu and collaborators found that the charge contained in white dwarfs affects their structure, they have larger masses and radii than the uncharged ones. In \cite{liu2014}, the electric charge distribution is considered such as the charge density is proportional to the energy density, $\rho_e=\alpha\rho$ ($\alpha$ being a proportional constant). The choice of this charge distribution is an appropriate assumption in the sense that a  star with a large mass could contain a large quantity of charge. Liu et al. found that the charged white dwarfs may attain values up to $3.0M_{\odot}$.

The study developed in \cite{liu2014} belongs to a large group of works where charged stars are investigated. Within them, we found some research where the influence of the electrical charge distribution at the stellar structure of polytropic stars \cite{raymalheirolemoszanchin,alz-poli-qbh,alz-rel-poli-qbh,azam2017,azam2016}, incompressible spherical objects \cite{defelice_yu,defelice_siming,annroth,alz-incom-qbh}, ani\-so\-tro\-pic stars \cite{Maurya2017} and strange stars \cite{negreiros2009,negreiros2011}, and in the radial stability against small perturbation  \cite{brillante2014,arbanil_malheiro2015} and in gravitational collapse \cite{ghezzi2005} are investigated. It is important to mention that, although charged stars are anisotropic objects, these maintain their spherical symmetry, since the energy momentum tensor components $T_{0}^{0}(r)$ and $T_{1}^{1}(r)$ are equals, being $r$ the radial coordinate (see, e.g., \cite{arbanil_malheiro2016,herrera1997,mak_harko2003}). In \cite{Maurya2015}, Maurya \textit{et al}. have obtained feasible stellar models of charged compact stars by assuming three different solutions for static spherically symmetric metric, and they have validated their model by comparing it with some observational parameters of compact objects, which are candidates to be strange stars.

In this work, we investigate the influence of the surface electric charge in the stellar structure and stability of white dwarfs. We consider that the electric charge at the star's surface follows a Gaussian distribution, such as is considered in \cite{negreiros2009,negreiros2011}. This consideration is realized since at  white dwarfs' atmospheres the electrons and ions could play an important role producing strong electric fields \cite{fassbinder1996}.
An important point that must be analyzed in charged stars is the Schwinger limit. It states that when electric fields  attains values around $\sim1.3\times10^{16}\,[\rm V/cm]$, these electric fields interact with the vacuum thus producing electron-positron pairs. Due to this phenomena, the electric fields decay to a value below the threshold \cite{schwinger1950}. It is important to point out that in neutron stars the magnitude of the electric field necessary to appreciate effects on the structure of these stars are at the order of $10^{19} \rm V/cm$ \citep{negreiros2009,raymalheirolemoszanchin}, i.e., much higher than Schwinger limit. This has been always advocated as a physical restriction to consider charge effects on the structure, at least, of neutron stars. Thus, in this article the Schwinger limit is analyzed.

The article is divided as follows. The stellar structure equations, the equation of state and the electric charge profile are described in section \ref{section_eos}. In section \ref{results} the stellar equilibrium configurations of charged white dwarfs are presented. The stability of these stars under small radial perturbations is investigated in \ref{results1}, using the results derived from the static method.  In section \ref{charge_radius} we discuss about the charge-radius relation found for the white dwarfs under analyze. Finally, we conclude in section \ref{conclusions}. Throughout the present work, we use geometric units.

\section{Stellar structure equation, equation of state and electric charge profile}\label{section_eos}

\subsection{Stellar structure equations}

The charged fluid contained in the spherical symmetric object is described by the energy-momentum tensor: 
\begin{equation}\label{TEM}
T_{\mu\nu}=(\rho+p)u_{\mu}u_{\nu}+pg_{\mu\nu}+\frac{1}{4\pi}\left(F^{\mu\gamma}F_{\varphi\gamma}-\frac{1}{4}g_{\mu\nu}F_{\gamma\beta}F^{\gamma\beta}\right),
\end{equation}
with $\rho$ and $p$ being the energy density and the pressure fluid, respectively. Moreover, $u_\mu$, $g_{\mu\nu}$ and $F^{\mu\gamma}$ stand the fluid's four velocity, the metric tensor and the Faraday-Maxwell tensor, respectively.

The interior background space-time of the star, in Schw\-arzschild coordinates, it is assumed of the following form
\begin{equation}\label{metric}
ds^2=-e^{\nu}dt^2+e^{\lambda}dr^2+r^2\left(d\theta^2+\sin^2\theta d\phi^2\right),
\end{equation}
where the functions $\nu=\nu(r)$ and $\lambda=\lambda(r)$ dependents on the radial coordinate $r$ only. 

The Maxwell-Einstein equations in such line element lead to the following set of stellar structure equations:
\begin{eqnarray}
&&\frac{dq}{dr}=4\pi\rho_{e}r^2e^{\lambda/2},\label{qo}\\
&&\frac{dm}{dr}=4\pi r^2\rho+\frac{q}{r}\frac{dq}{dr},\label{mo}\\
&&\frac{dp}{dr}=-(p+\rho)\left(4\pi rp+\frac{m}{r^2}-\frac{q^2}{r^3}\right)e^{\lambda}+\frac{q}{4\pi r^4}\frac{dq}{dr},\label{tov}\\
&&\frac{d\nu}{dr}=-\frac{2}{(p+\rho)}\frac{dp}{dr}+\frac{2q}{4\pi r^4(p+\rho)}\frac{dq}{dr},\label{nuo}
\end{eqnarray}
where the metric component $e^{\lambda}$ is given by the equality:
\begin{equation}\label{lambdao}
e^{\lambda}=\left(1-\frac{2m}{r}+\frac{q^2}{r^2}\right)^{-1}.
\end{equation}
The variables $q$, $m$ and $\rho_e$ showed before represent respectively the electric charge, the mass within the sphere of radius $r$, and the charge density. Equation \eqref{tov} gives the Tolman-Oppenheimer-Volkoff equation \cite{tolman,oppievolkoff} modified to the inclusion of the electric charge \cite{bekenstein}.

\subsection{Equation of state}

It is considered that the pressure and the energy density of the fluid contained in the spherical object follow the relations:
%
\begin{eqnarray}
p(k_F)&=&\frac{1}{3\pi^2\hbar^3}\int_0^{k_F}\frac{k^4}{\sqrt{k^2+m_e^2}}dk,\label{eos}
\\
\rho(k_F)&=&\frac{1}{\pi^2\hbar^3}\int_0^{k_F}\sqrt{k^2+m_e^2}k^2dk+\frac{m_N\mu_e}{3\pi^2 \hbar^3}k_F^3 ,\label{energy} 
\end{eqnarray}
where $m_e$ is the electron mass, $m_N$ represents the nucleon mass, $\hbar$ denotes the reduced Planck constant, $\mu_e$ states the ratio between the nucleon number and atomic number for ions and $k_F$ depicts the Fermi momentum of the electron, review \cite{chandra,shapiro}. Eq.~\eqref{eos} is the electric degeneracy pressure and the first and second terms of the right hand side of Eq.~\eqref{energy} are respectively the electron energy density and the energy density related to the rest mass of the nucleons.

It is important to mention that for numerical and analytical analysis the equations \eqref{eos} and \eqref{energy} are using in the form
\begin{eqnarray}\label{eossimplified}
p(x) &=& \epsilon_0 f(x),\\
\rho(x) &=& \epsilon_0 g(x),
\end{eqnarray}
where
\begin{eqnarray}
f(x)&=&\frac{1}{24}\left[(2x^3-3x)\sqrt{x^2+1}+3\textrm{asinh} x\right]\\
g(x)&=&\frac{1}{8}\left[(2x^3+x)\sqrt{x^2+1}-\textrm{asinh} x\right]\nonumber\\ &&+1215.26x^3,
\end{eqnarray}
being $\epsilon_0=m_e/\pi^2\lambda_e^3$ and $x=k_F/m_{e}$ the dimensionless Fermi momentum, with $\lambda_e$ representing the electron Compton wavelength. In the last two equations, it is taking into account $\mu_e=2$.

\subsection{Electric charge profile}

We assume that the interior and the atmosphere of the white dwarfs are made of respectively by degenerate and nondegenerate
matter. This hypotheses is consistent since the energy density decays toward the white dwarf' surface, where it drops to zero. At this point
the white dwarf must have a nondegenerate atmosphere (review, e.g., \cite{goldreich1969,koester1990}). For the white dwarfs under analyzes, we consider that the majority amount of charged is concentrated at the surface of star. As aforementioned, the surface electric charge of the white dwarf could develop an important task, producing superficial strong electric fields \cite{fassbinder1996}. Thus, following \cite{negreiros2009}, we model the distribution of the electric charge in terms of a Gaussian, hence the charge distribution can be regarded as the following ansatz:
\begin{equation}\label{densitycharge}
\rho_{e}=k\exp\left[-\frac{(r-R)^2}{b^2}\right],
\end{equation}
being $R$ the radius of the star in the uncharged case. The constant $b$ is the width of the electric charge distribution, in this case, we consider $b=10\,[\rm km]$. For small width of this layer, the stellar structure of the white dwarf does not change significantly. 

In order to determine the constant $k$, we require to the equality:
\begin{equation}\label{kapa}
\sigma=\int_{0}^{\infty}4\pi r^2\rho_{e}dr,
\end{equation}
with $\sigma$ being the magnitude proportional to the electric char\-ge distribution. $\sigma$ would represents the total charge of the star if we was working out in flat space-time. Using Eq.~\eqref{densitycharge} in Eq.~\eqref{kapa}, it yields: 
\begin{equation}\label{kapa2}
4\pi k=\sigma\left(\frac{\sqrt{\pi}bR^2}{2}+\frac{\sqrt{\pi}b^3}{4}\right)^{-1},
\end{equation}
this last equation connects the constant $k$ and $\sigma$.

\subsection{Numerical method and boundary conditions}

The stellar structure equilibrium equations \eqref{qo}, \eqref{mo}, \eqref{tov} and \eqref{lambdao}, together with the equations \eqref{eos} and \eqref{densitycharge}, are integrated using the Runge-Kutta fourth order method, for different values of $\sigma$ and $\rho_c$. These equations are integrated from the center $r=0$ towards the surface of the star $r=R$. The integration of the aforementioned equations begins with the values in the center:
\begin{eqnarray}
m(0)&=&0,\;\;q(0)=0,\;\;\rho(0)=\rho_c,\nonumber
\\
p(0)&=&p_c,\;\;\;\;\;{\rm and}\;\;\;\;\;\rho_e(0)=\rho_{ec}.
\end{eqnarray}
The integration of the stellar structure equations ends when the star's surface is attained, $p(R)=0$. At the surface of the object the interior solution matches smoothly with the vacuum exterior Reissner-Nordstr\"om line element. This indicates that the inner and outer metric function are related by the equality
\begin{equation}
e^{\nu(R)}=e^{-\lambda(R)}=1-\frac{2M}{R}-\frac{Q^2}{R^2},
\end{equation}
where $M$ and $Q$ being the total mass and total charge of the sphere.

\section{Stellar equilibrium configurations of charged white dwarfs}\label{results}

\begin{figure}[ht]
\begin{center}
\includegraphics[width=0.97\linewidth]{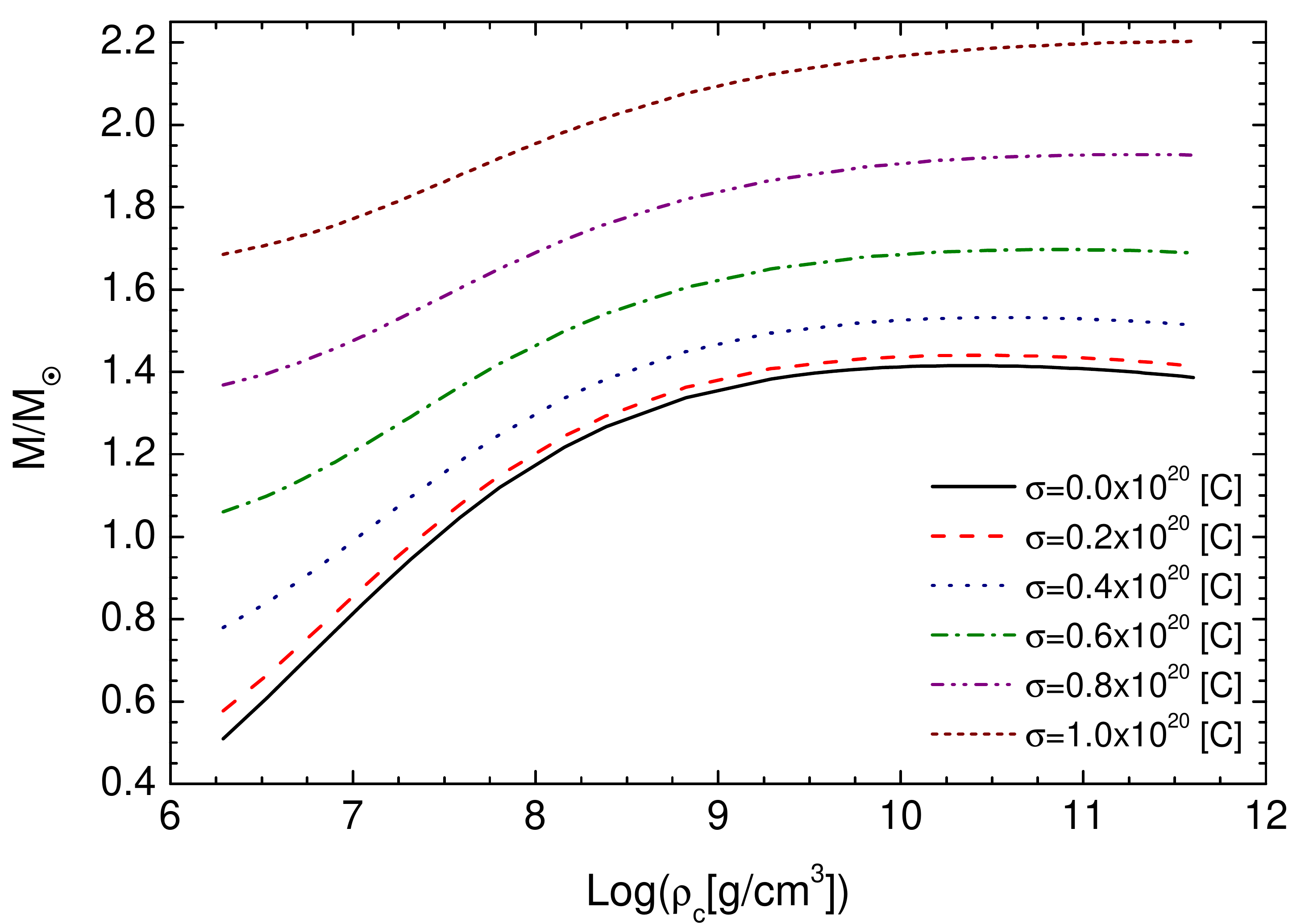}
\caption{Total mass, in Solar masses $M_{\odot}$, versus central mass density of the star for six values of $\sigma$. The unit for the constant $\sigma$ is $[\rm C]$.}
\label{chWD_MRHO}
\end{center}
\end{figure}

Fig. \ref{chWD_MRHO} shows the behavior of the total mass $M/M_{\odot}$ with the central energy density, for six values of $\sigma$. The considered central energy densities are in the interval $2\times10^{6}[\rm g/cm^3]$ to $4\times10^{11}[\rm g/cm^3]$. The lower limit $\rho_c=2\times 10^{6}[\rm g/cm^3]$ is the mean density for white dwarfs and in the upper limit $\rho_c=4\times 10^{11}[\rm g/cm^3]$ the neutron drip limit is reached, i.e., in this point, the white dwarf turn into a neutron star. In the cases where $\sigma\leq0.8\times10^{20}[\rm C]$, we note that the total mass grows with the central energy density until attain a maximum mass point, after that point, the mass starts to decrease with the grows of the density of energy center. In turn, in the case $\sigma=1.0\times10^{20}[\rm C]$ the mass grows monotonically with the central energy density, i.e., we do not found a turning point.

\begin{figure}[ht]
\begin{center}
\includegraphics[width=0.97\linewidth]{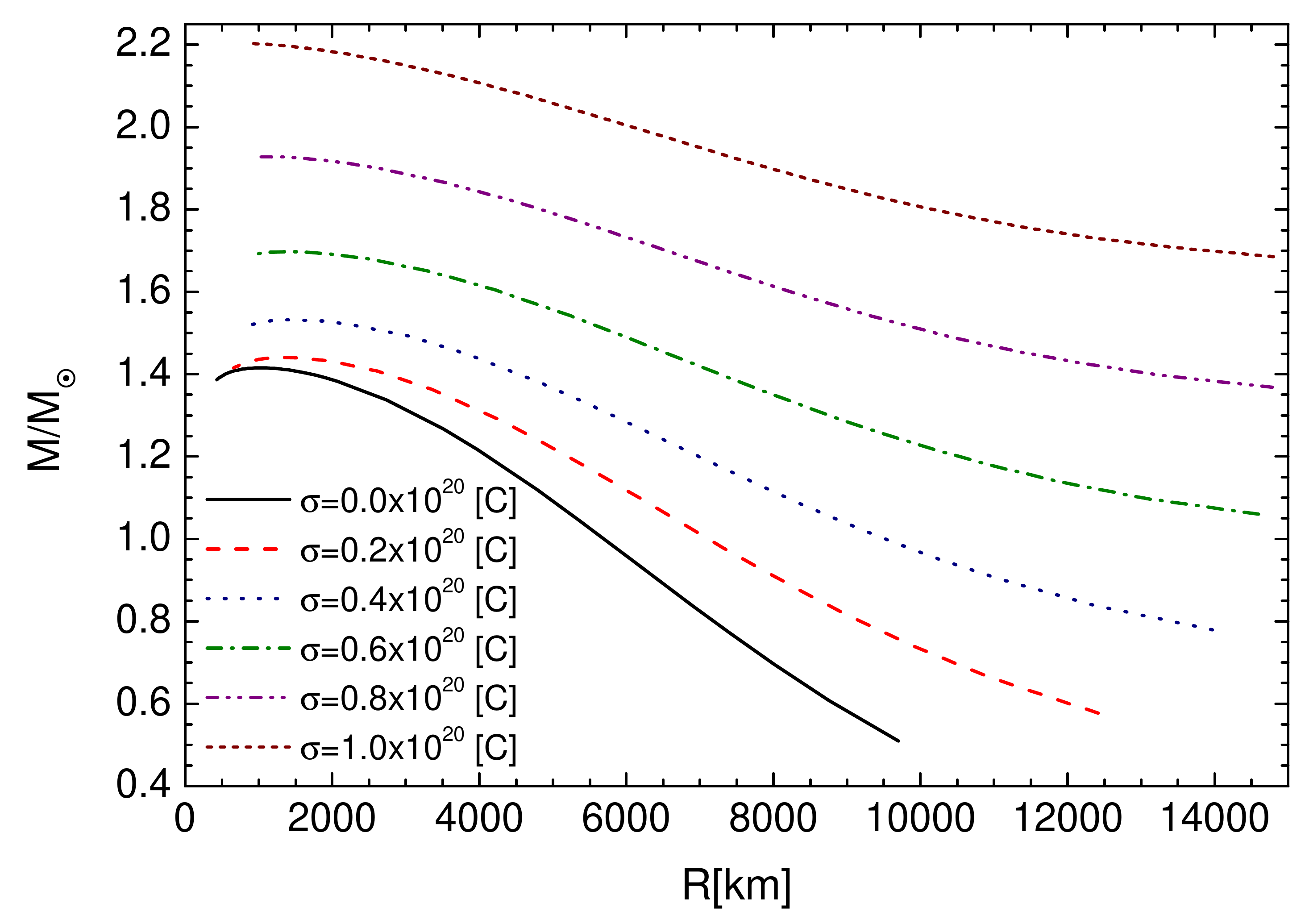}
\caption{The radius of the charged white dwarf as a function of the mass for different values of $\sigma$. }
\label{chWD_MR}
\end{center}
\end{figure}

On the other hand, we observe that exist a dependency of the total mass with the parameter $\sigma$. For a larger $\sigma$ more massive stars are obtained. As can be noted in Fig.~\ref{chWD_MRHO}, we obtain more massive white dwarfs using $\sigma=1.0\times10^{20}[\rm C]$. In this case, the mass whose respective electric field saturates the Schwinger limit ($\sim1.3\times10^{16} [\rm V/cm]$) is  $2.199M_{\odot}$, this is attained in the central energy density $1.665\times 10^{11}[\rm g/cm^3]$. This mass is between the masses estimated for the super-Chandrasekhar white dwarfs, $2.1$-$2.8M_{\odot}$ \cite{taub2011,silverman2011}. From this, we understand that a surface electric field produces considerable effects in the masses of white dwarfs. In addition, it is important to mention that the grow of the mass with $\sigma$ can be understood since $\sigma$ is related with the total charge contained in the star. The charge produces a force which helps to the one generated by the radial pressure to support more mass against the gravitational collapse.

\begin{table*}[h]
\centering
\begin{tabular}{cccccc}
\hline \hline\
$\sigma[\rm C]$ & $M/M_{\odot}$ & $R[\rm km]$ & $\rho_c$[\rm g/cm$^3$] & $Q[\rm C]$ & $E[\rm V/cm]$\\
\hline\
$0.0\times10^{20}$ & $1.416$ & $1021$ & $2.307\times 10^{10}$ & - & -\\
$0.2\times10^{20}$ & $1.441$ & $1299$ & $3.043\times 10^{10}$ & $4.055\times 10^{19}$ & $2.158\times 10^{15}$\\
$0.4\times10^{20}$ & $1.532$ & $1539$ & $3.456\times 10^{10}$ & $8.109\times 10^{19}$ & $3.078\times 10^{15}$\\
$0.6\times10^{20}$ & $1.698$ & $1336$ & $6.613\times 10^{10}$ & $1.222\times 10^{20}$ & $6.149\times 10^{15}$\\
$0.8\times10^{20}$ & $1.928$ & $1166$ & $1.942\times 10^{11}$ & $1.637\times 10^{20}$ & $1.081\times 10^{16}$\\
$1.0\times10^{20}$ & $2.203$ & $916.8$ & $4.000\times 10^{11}$ & $2.058\times 10^{20}$ & $2.200\times 10^{16}$\\
\hline
\hline
\end{tabular}\\
\label{table_maximum_masses}
\caption{The constant $\sigma$ and the maximum masses of the electrically charged white dwarfs with their respective radii, central densities, charges and electric fields at the surface of the stars.}
\end{table*}

In Fig. \ref{chWD_MR} the total mass as a function of the radius for few values of $\sigma$ is observed. In the uncharged case $\sigma=0.0$ the curve is close to attain the typical Chandrasekhar limit, $1.44M_{\odot}$ \cite{Chandrasekhar1931,Chandrasekhar1935}, however, in the charged case $\sigma\neq0$ we found masses that overcome this typical limit. For instance, in the case $\sigma=1.0\times10^{20}[\rm C]$ the mass whose electric field saturates the Schwinger limit is around $2.199M_{\odot}$. Again, we mention that high values of white dwarf masses (around the super-Chandrasekhar white dwarf masses) can be achieved taking into account a surface electrical charge at the white dwarf.

\begin{figure}[ht]
\begin{center}
\includegraphics[width=0.97\linewidth]{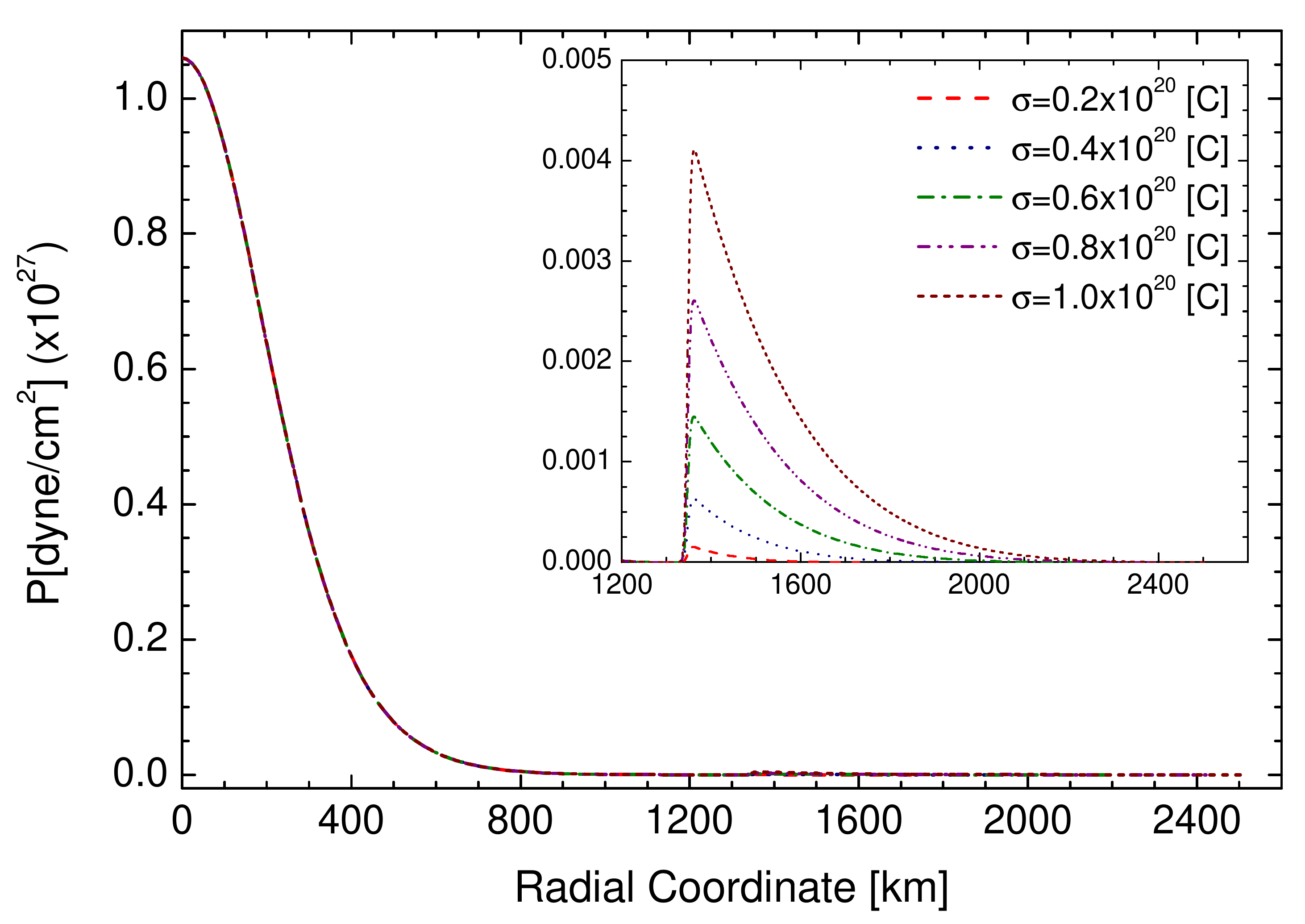}
\caption{Profile of the pressure inside the star as a function of the radial pressure for five different values of $\sigma$ and for $\rho_c=10^{10}[\rm g/cm^3]$.}
\label{pressure}
\end{center}
\end{figure}

With the purpose to observe that the electric charge is only distributed near the star's surface, the pressure profile inside the white dwarf as a function of the radial coordinate is plotted in Fig.~\ref{pressure}, where few values of $\sigma$ and $\rho_c=10^{10}[\rm g/cm^3]$ are considered. In figure we can note that the pressure decays monotonically toward the baryonic surface, when this is attained the pressure grows abruptly due to the beginning of the electrostatic layer, after this point the pressure decrease with the radial coordinate until attain the star's surface ($P=0$). Thus, through this result, we can clearly note that the electric charge is distributed as a spherical shell close to the surface of the white dwarf. 

\begin{figure}[ht]
\begin{center}
\includegraphics[width=0.97\linewidth]{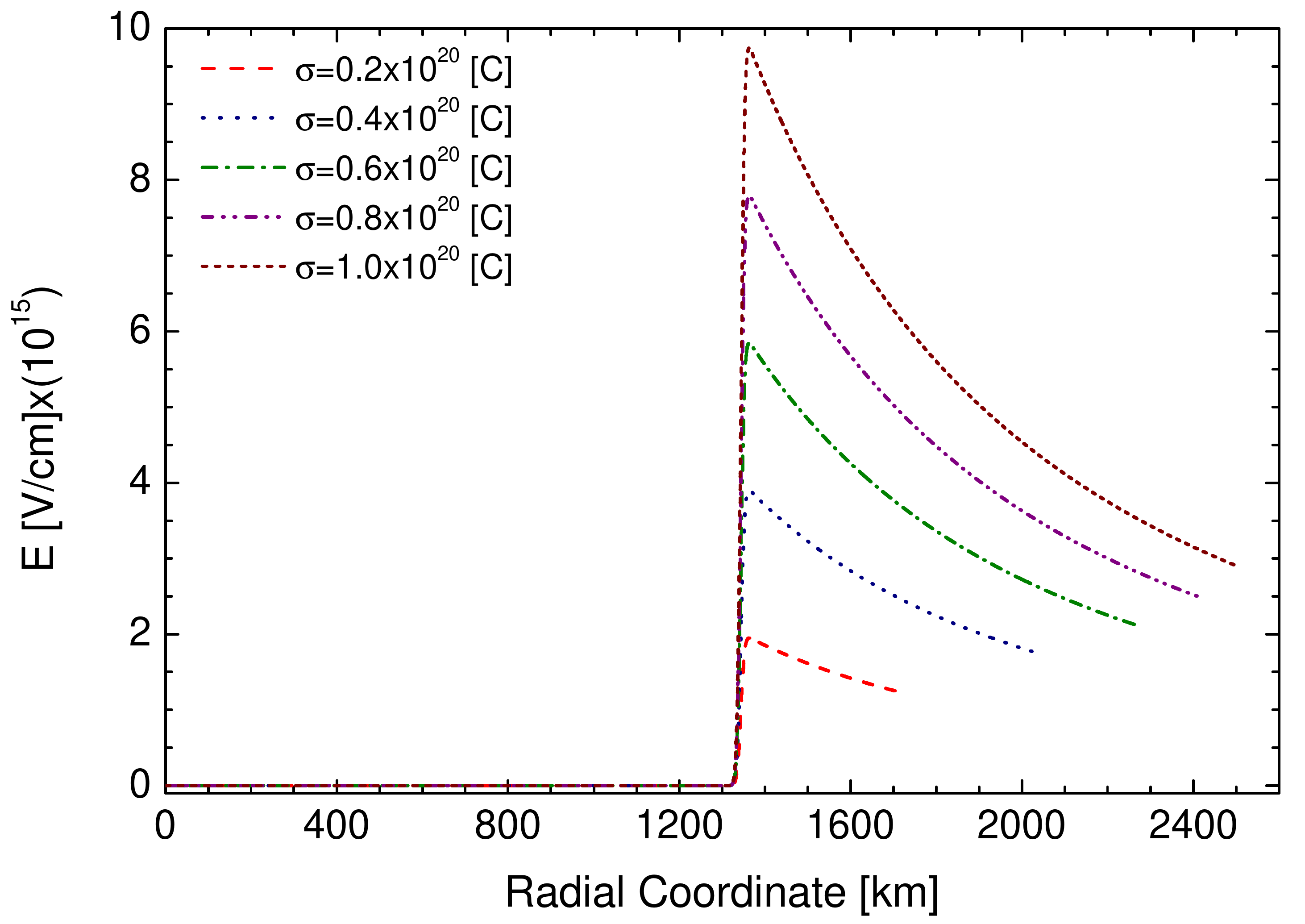}
\caption{Electric field as a function of the radial coordinate inside the charged white dwarf, for five values of $\sigma$ and one of $\rho_c$. The central energy density $10^{10}[\rm g/cm^3]$ is considered.}
\label{efield}
\end{center}
\end{figure}

The behavior of the electric field in the star is showed in Fig. \ref{efield}. On figure is employed five different values of $\sigma$ and $\rho_c=10^{10}[\rm g/cm^3]$. As can be seen, in each case presented, the electric field exhibit a very abrupt increase from zero to $10^{15-16}[\rm V/cm]$, thus indicating that the baryonic surface ends and starts the electrostatic layer. Once the electric surface is distributed like a thin layer close to the surface, the electric field sharply weaken with the grow of the radial coordinate such as is showed in figure. 

It is important to emphasize that the electric field could be reduced, once taking into account the change that the electric potential screening may suffer with the increment of the radial distance (see \cite{Akbari2014}). The analysis of such a situation is left for future investigation.

It is worth mentioning that the electric field found in the charged white dwarfs cases are $10^{4}$ times lower than those found in charged strange stars ones (see, e.g., \cite{negreiros2009,negreiros2011,arbanil_malheiro2015}). This can be understand since white dwarfs have very larger radii than the strange stars.

The values for $\sigma$ employed, the maximum mass values found in the range of central energy densities considered, with their respective central energy densities, total radius, total charge and electric field are shown in Table I.

\section{About the radial stability of charged white dwarf}\label{results1}

\begin{figure}[ht]
\begin{center}
\includegraphics[width=0.97\linewidth]{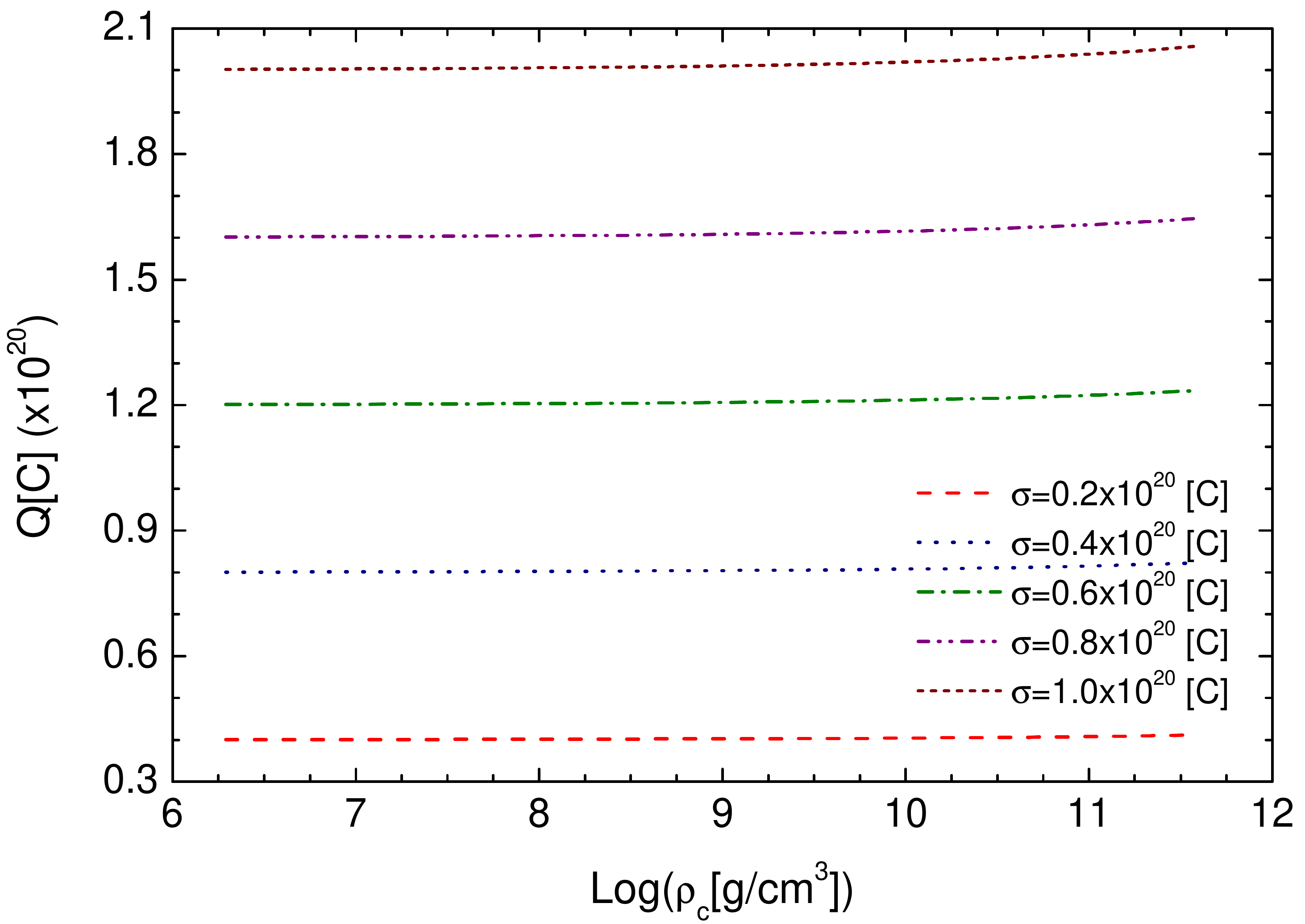}
\caption{Total electric charge against the central energy density for some values of $\sigma$.}
\label{Q_rhoc_sigma}
\end{center}
\end{figure}

Inspired in the study of the turning-point method for axisymmetric stability of uniformly rotating relativistic stars with the angular momentum fixed \cite{friedman1988} (see also \cite{rezzolla2011} for a detailed discussion about that theme), in \cite{arbanil_malheiro2015} is shown that the stability of charged objects could be investigated using the results obtained from the hydrostatic equilibrium equation. The authors in \cite{arbanil_malheiro2015} found that along a sequence of charged stars with increasing central energy density and with fixed total charge, the maximum mass equilibrium configuration states the onset of instability. Thus, likewise, in this section we use the results derived from the equilibrium configurations to determine the maximum mass point which marks the start of the instability.

The total charge of white dwarf as a function of the central energy density is plotted in Fig.~\ref{Q_rhoc_sigma}, taking into account five different values of $\sigma$. In figure we can see that along the sequence of equilibrium configurations with increasing central energy density the total charge is nearly constant. In this case, we understand that the maximum mass point must marks the onset of instability. In other words, the regions made of stable and unstable charged white dwarfs shall be distinguished through the relations $dM/d\rho_c>0$ and $dM/d\rho_c<0$, respectively.

Additionally, in Fig.~\ref{Q_rhoc_sigma}, it can be observed also that the electrical charge that produce considerable effects in the structure of white dwarfs is around $~10^{20}[\rm C]$. This amount of charge is similar to those found in the studies, for instance, of polytropic stars \cite{raymalheirolemoszanchin,alz-poli-qbh}, incompressible stars \cite{defelice_yu,defelice_siming,annroth,alz-incom-qbh}, strange stars \cite{arbanil_malheiro2015,negreiros2009,negreiros2011} and white dwarfs \cite{liu2014}, where, certainly, the electric charge is considered.

\section{Universal charge-radius relation and maximum total charge of white dwarfs}\label{charge_radius}

In \cite{madsen2008}, Madsen demonstrates that the electrical charge of a spherically symmetrical static object is limited by the creation of electron-positron pairs in super critical electric fields. Taking into account a timescale $\tau<<\infty$, the net positive charge $Q$ is directly proportional to the square of the star's radius $R_{\rm km}$, i.e.,:
\begin{equation}\label{madsen_equation}
Q=\beta e R^{2}_{\rm km},
\end{equation} 
with $\beta$ and $e$ being the proportionality constant and charge of a proton, respectively. The proportionality constant $\beta$ is directly related with the timescale chosen $\tau$, for lower time\-scale a larger $\beta$ is derived. Madsen found that for a timescale $\tau=1.0\,s$ is obtained $\beta=7.0\times 10^{31}$, and for $\tau=1.0\times 10^{-10}\,s$, i.e., a typical weak interaction timescale, $\beta=1.68\times 10^{32}$. 

For white dwarfs we can consider the electromagnetic interaction, consequently,  the timescale can be regarded to be around $10^{-18}\,s$, thereby the constant $\beta$ becomes equal to $3.34\times 10^{33}$. In this case, we obtain that the more massive white dwarf (see Table I) would allow the maximum charge:
\begin{equation}\label{Qmax}
Q_{{\rm max}}\approx 5.0\times 10^{20}\,[\rm C].
\end{equation}
Thus, the quantity of charge found in the most massive white dwarf ($Q=2.045\times10^{20}[\rm C]$) is under the maximum charge limit of Eq.~\eqref{Qmax}.

\section{Conclusions}\label{conclusions}

The static equilibrium configurations and the stellar radial stability of charged white dwarfs are investigated in this work. Both studies are analyzed through the results derived from the hydrostatic equilibrium equation, the Tolman-O\-ppen\-hei\-mer-Volkoff equation, modified to include the electrical part. For the interior of white dwarfs, we consider that the equation of state follows the employed  for the fully degenerated electron gas \cite{chandra,shapiro}. In addition, we assume a Gaussian distribution of charge of $\sim 10\,[\rm km]$ thickness close the star's surface. It is important to mention that the interior solution match smoothly to the exterior Reissner-Nordstr\"om vacuum solution.

We observe that for larger total charge, more massive stellar objects are found. For instance, the increment of the total charge from $0$ to $2.058\times10^{20} [{\rm C}]$ allows to increase the total mass in approximately $55.58\%$, growing from $1.416\,M_{\odot}$ to $2.203\,M_{\odot}$. This increment in the mass of the star is explained since the electric charge acts as an effective pressure, thus helping the hydrodynamic pressure to support more mass against the gravitational collapse.  It is worth mentioning that for the total  electric charge $2.058\times10^{20} [{\rm C}]$, we found that the Schwinger limit is saturated for a white dwarf with $\sim2.2\,M_{\odot}$. This total mass is within the interval of white dwarf considered as super-Chandrasekhar white dwarfs \cite{taub2011,silverman2011}. From the aforementioned, we can understand that a surface distribution of charge could plays an important role in the existence of the super-Chandrasekhar white dwarfs. 

On the other hand, the stability against small radial perturbations of charged white dwarfs are analyzed using a sequence of equilibrium configurations with increasing central energy density, where these spherical objects are constituted by an equal total electric charge. In this types of sequences, the maximum mass point marks the onset of the instability, see \cite{arbanil_malheiro2015}. From this, we can say that the regions where lay stable and unstable white dwarfs can be distinguished by the inequalities $dM/d\rho_c>0$ and $dM/d\rho_c<0$, respectively. 

\begin{acknowledgements}
Authors would like to thank Coordena\c{c}\~ao de A\-per\-fei\c{c}oamento de Pessoal de N\'ivel Superior-CAPES and Funda\c{c}\~ao de Amparo {\`a} Pesquisa do Estado de S\~ao Paulo-FAPESP, under the thematic project 2013/26258-4, for the financial supports. GAC also thanks to Professors Dr. P. J. Pompeia and Dr. P. H. R. S. Moraes for discussion and support under the preparation of this work.

\end{acknowledgements}

\end{document}